\documentclass[aip,apl,amsmath,amssymb,reprint]{revtex4-1}
\usepackage{graphicx}
\usepackage{dcolumn}
\usepackage{bm}
\usepackage{mathrsfs} 
\usepackage{color}

\usepackage{siunitx}
\newcommand{\um}{\SI{}{\micro\meter}}
\newcommand{\uW}{\SI{}{\micro\watt}}
\newcommand{\Xv}{$\langle 100 \rangle$}
\newcommand{\Lv}{$\langle 111 \rangle$}

\begin{document}

\preprint{AIP/123-QED}

\title[Charge Transport in Silicon]{Spatial Imaging of Charge Transport in Silicon at Low Temperature}

\affiliation{Department of Physics, Stanford University, Stanford, CA 94305 USA}
\affiliation{SLAC National Accelerator Laboratory/Kavli Institute for Particle Astrophysics and Cosmology, 2575 Sand Hill Road, Menlo Park, CA 94025 USA}
\affiliation{Fermi National Accelerator Laboratory, Center for Particle Astrophysics, Batavia, IL 60510 USA}
\affiliation{Department of Physics, Santa Clara University, Santa Clara, CA 95053 USA}
\affiliation{Department of Physics, University of Illinois at Urbana-Champaign, Champaign, IL 61820 USA}
\affiliation{Department of Physics, San Diego State University, San Diego, CA 92182 USA}

\author{R.A.~Moffatt} \affiliation{Department of Physics, Stanford University, Stanford, CA 94305 USA}

\author{N.A.~Kurinsky}\email{kurinsky@fnal.gov} \affiliation{Department of Physics, Stanford University, Stanford, CA 94305 USA} \affiliation{SLAC National Accelerator Laboratory/Kavli Institute for Particle Astrophysics and Cosmology, 2575 Sand Hill Road, Menlo Park, CA 94025 USA} \affiliation{Fermi National Accelerator Laboratory, Center for Particle Astrophysics, Batavia, IL 60510 USA}

\author{C.~Stanford} \affiliation{Department of Physics, Stanford University, Stanford, CA 94305 USA}

\author{J.~Allen} \affiliation{Department of Physics, Stanford University, Stanford, CA 94305 USA}\affiliation{Department of Physics, University of Illinois at Urbana-Champaign, Champaign, IL 61820 USA}

\author{P.L.~Brink} \affiliation{SLAC National Accelerator Laboratory/Kavli Institute for Particle Astrophysics and Cosmology, 2575 Sand Hill Road, Menlo Park, CA 94025 USA}

\author{B.~Cabrera}\email{cabrera@stanford.edu} \affiliation{Department of Physics, Stanford University, Stanford, CA 94305 USA} \affiliation{SLAC National Accelerator Laboratory/Kavli Institute for Particle Astrophysics and Cosmology, 2575 Sand Hill Road, Menlo Park, CA 94025 USA}

\author{M.~Cherry} \affiliation{SLAC National Accelerator Laboratory/Kavli Institute for Particle Astrophysics and Cosmology, 2575 Sand Hill Road, Menlo Park, CA 94025 USA}

\author{F.~Insulla} \affiliation{Department of Physics, Stanford University, Stanford, CA 94305 USA}

\author{F.~Ponce} \affiliation{Department of Physics, Stanford University, Stanford, CA 94305 USA}

\author{K.~Sundqvist} \affiliation{Department of Physics, San Diego State University, San Diego, CA 92182 USA}

\author{S.~Yellin} \affiliation{Department of Physics, Stanford University, Stanford, CA 94305 USA}

\author{J.J.~Yen} \affiliation{Department of Physics, Stanford University, Stanford, CA 94305 USA}

\author{B.A.~Young} \affiliation{Department of Physics, Santa Clara University, Santa Clara, CA 95053 USA}

\date{\today}

\begin{abstract}
We present direct imaging measurements of charge transport across a 1\,cm$\times$1\,cm$\times$4\,mm crystal of high purity silicon ($\sim$20 k$\Omega$cm) at temperatures between 500\,mK and and 5\,K. We use these data to determine the intervalley scattering rate of electrons as a function of the electric field applied along the $\langle 111 \rangle$ crystal axis, and we present a phenomenological model of intervalley scattering that explains the constant scattering rate seen at low-voltage for cryogenic temperatures. We also demonstrate direct imaging measurements of effective hole mass anisotropy, which is strongly dependent on both temperature and electric field strength. The observed effects can be explained by a warping of the valence bands for carrier energies near the spin-orbit splitting energy in silicon. 
\end{abstract}

\maketitle

\section{Introduction}



\begin{figure}[t]
\centering
\includegraphics[width=0.23\textwidth]{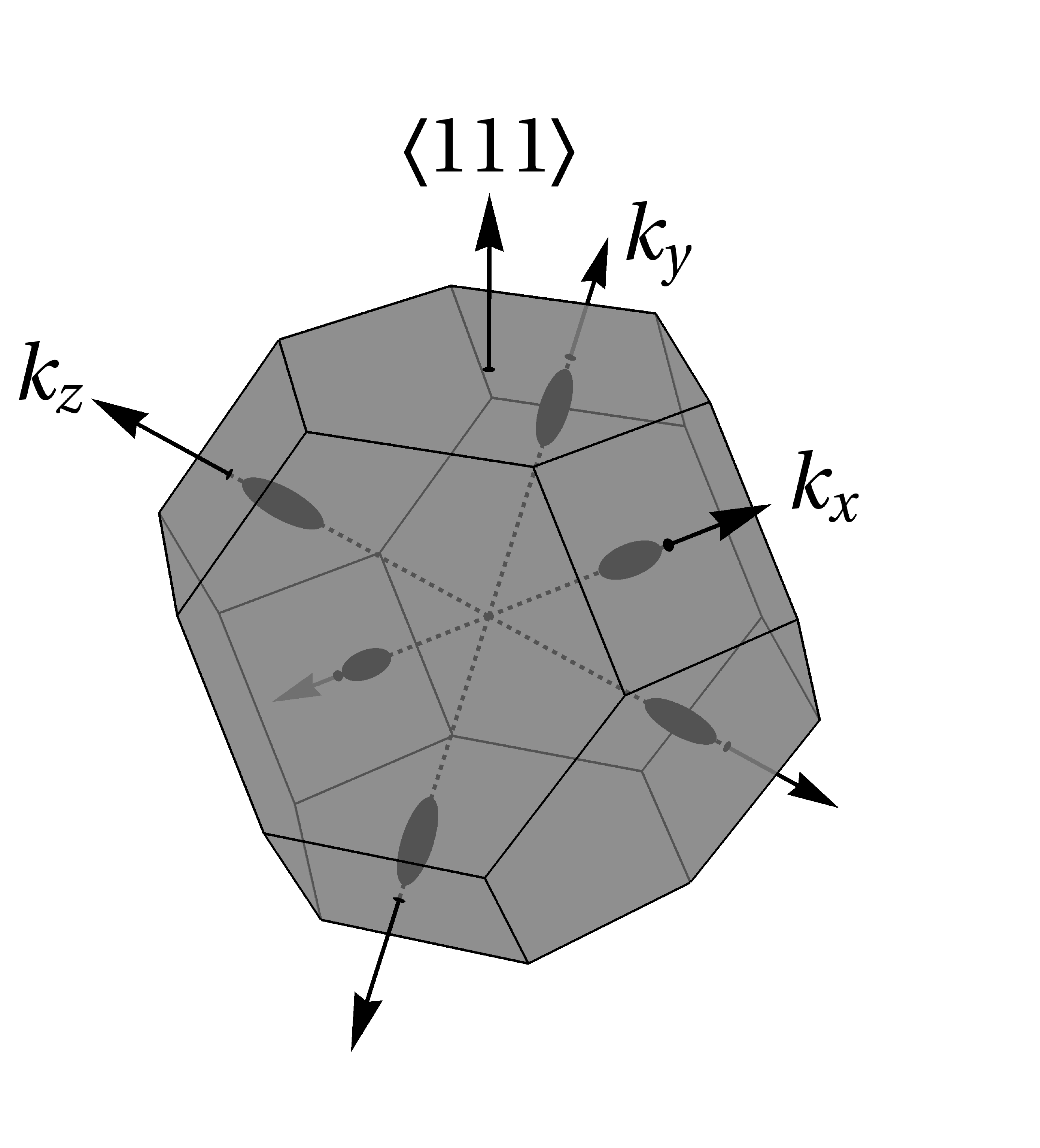}
\includegraphics[width=0.23\textwidth]{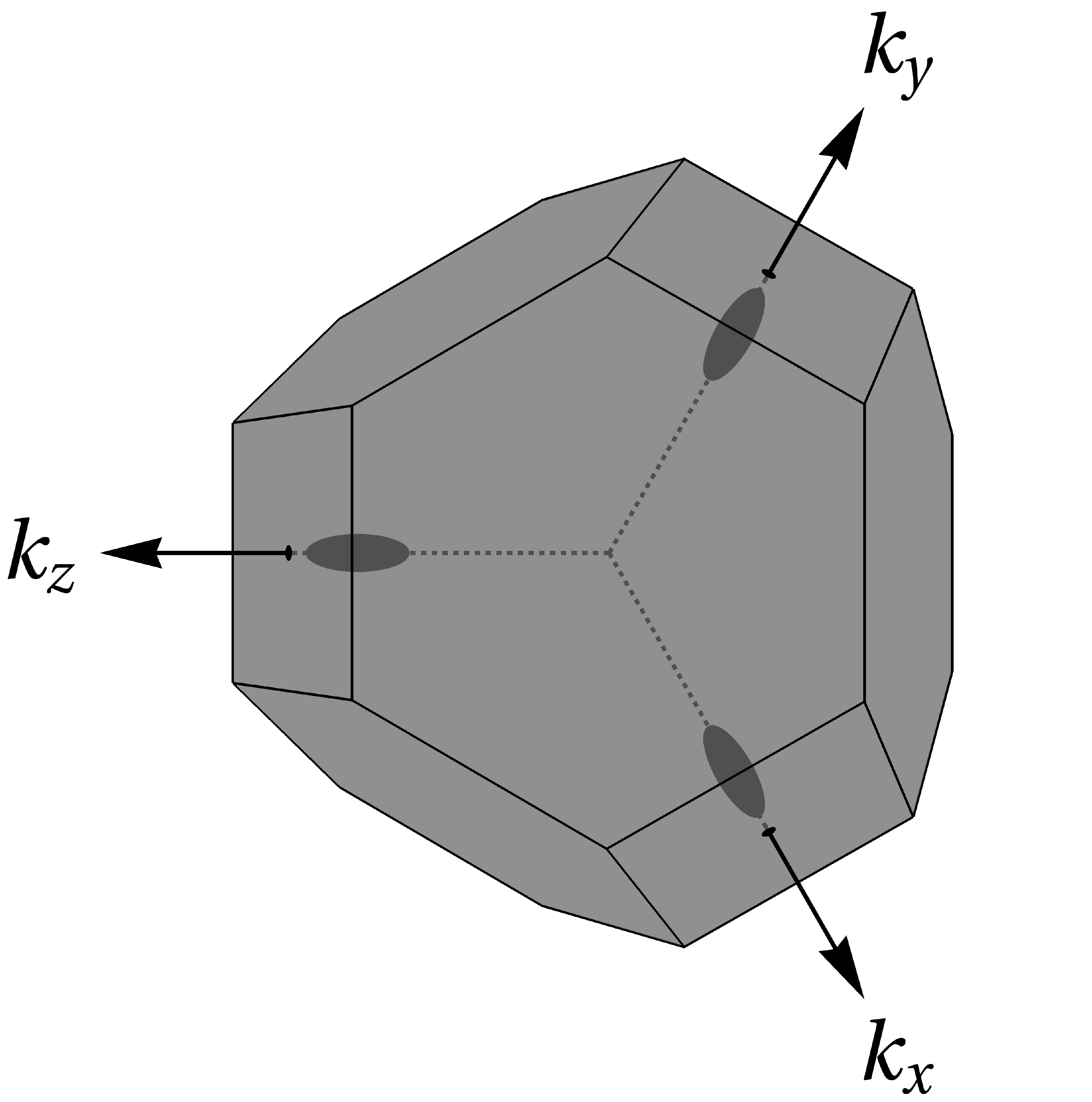}
\caption{\textbf{Left}: The Brillouin zone for a face-centered cubic lattice orientated with the z-axis along the \Lv\ direction. In Si, the electron energy minima, or valleys, are located near the edge of the Brillouin zone in the $k_x$, $k_y$, and $k_z$ directions (black ovals). \textbf{Right}: The top half of the Brillouin zone viewed from the top face. When an electric field is applied to the crystal to drift the electrons upward, we expect the electrons to propagate along these three valleys and produce a triangular charge distribution pattern.}
\label{fig:brillouin}
\end{figure}

Silicon (Si) is an indirect band gap semiconductor with electron energy minima (valleys) displaced from zero momentum in the Brillouin zone along the \Xv\ vectors. Figure~\ref{fig:brillouin} shows surfaces of constant energy about these minima in the Brillouin zone, viewed off-axis as well as along the \Lv\ direction. These valleys are highly anisotropic; electrons have effective masses of 0.98~$m_{e}$ and 0.19~$m_{e}$ in the momentum space directions parallel and perpendicular to the valley axis, respectively\cite{Jacoboni}. If a small electric field is applied along the \Lv\ direction, acoustic scattering within these valleys will keep free electrons aligned with the valley axis, and produce a drift velocity which is aligned with that axis rather than the electric field. 
This will produce three spatially distinct charge concentrations, as illustrated in the right panel of Fig.~\ref{fig:brillouin}. 

As the electric field is increased, the mean carrier energy also increases, increasing the probability that a charge moving through the lattice will emit a high energy optical phonon and transition between Brillouin zone minima\cite{Jacoboni,ridley1999,SundqvistThesis,canali}, producing an intervalley scattering event. This will allow the spatially distinct minima to start to converge along the electric field lines through increased intervalley scattering. For moderate rates, where each charge undergoes at most a single intervalley transition, this manifests as a line connecting the minima, forming a triangle. For high intervalley scattering rates, charges effectively propagate along the electric field. 


In this paper, we present direct measurements of the intervalley scattering rate in Si through the imaging of the spatial distribution of electrons, and discuss the implications of this measurement for models of intervalley scattering at low applied E-field strength. In particular, we discuss an interpretation of a constant low electric field scattering rate as being due to neutral impurities. This experiment uses cryogenic Si crystals in the same operating conditions as our SuperCDMS dark matter detectors\cite{MoffattThesis,SundqvistThesis,kurinsky}, allowing us to refine models of athermal charge and phonon dynamics relevant to the operation of these detectors, and to validate our detector Monte Carlo.


An additional curiosity discussed in this paper is the anisotropic propagation of holes in silicon near our temperatures of interest ($\lesssim\SI{5}{\kelvin}$). Cyclotron resonance measurements showed in 1955 that holes have an anisotropic dispersion relation\cite{DKK}, 
and subsequent work demonstrated that this anisotropy becomes most pronounced when mean carrier energies are of the same order as the spin-orbit splitting energy\cite{kane,cardona}, which breaks the degeneracy between the heavy and light hole bands. The spin-orbit coupling energy in Si is $\sim 44$~meV\cite{cardona,ottaviani}, which is comparable to the mean hole energy induced by electric field heating at our operating temperatures, as expected from Monte Carlo simulations\cite{SundqvistThesis}. We present the direct imaging of this anisotropy, which was not seen in our previous result\cite{Moffatt} with Ge because its higher spin-orbit splitting energy ($\sim 200$~meV) makes it less pronounced under our operation conditions.

\section{Experimental Setup}\label{sec:exp}

The crystal under test was cut from a 4\,mm thick wafer of undoped ultra-high-purity float-zone silicon ($\sim$15 k$\Omega$-cm). The residual impurity ($10^{12}$\,cm$^{-3}$) was measured to be p-type. The front and back faces of the crystal are 1\,cm$\times$1\,cm, and lie in the \Lv\ plane. The front face is patterned with an aluminum-tungsten mesh electrode, with 20\% coverage~\cite{electrode}. The back face features a small, inner electrode in the center of the face, circular in shape with a diameter of 160\,\um, separated by a 10\,\um\ gap from the outer electrode, which covers the rest of the face. These electrodes are separately contacted, but are held at the same voltage to provide a near-uniform field across the crystal.

Free electrons and holes are produced in the crystal by exposing the front face to a 50\,ns pulse of 200\,\uW, 650\,nm laser light, focused to a 60\,\um\ diameter spot. Then, depending on the sign of the bias voltage, either the electrons or the holes propagate through the crystal and produce voltage pulses in the electrodes on the back face. In order to reduce scattering by background phonons, the crystal is cooled in a He-3 cryostat with a base temperature of 500\,mK.

The location of the laser spot on the front face can be controlled by a Micro-Electromechanical Systems (MEMS) mirror~\cite{mirrorcle}. Since scanning the laser across the front face and observing the charge collected on the fixed inner electrode is equivalent to having a fixed laser spot and a moving electrode, we are able to produce full, 2-D mappings of the charge density pattern. More details of the setup, including a diagram and picture, can be found in \citet{Moffatt}.

\begin{figure}[t]
    \centering
	\includegraphics[width=0.48\textwidth]{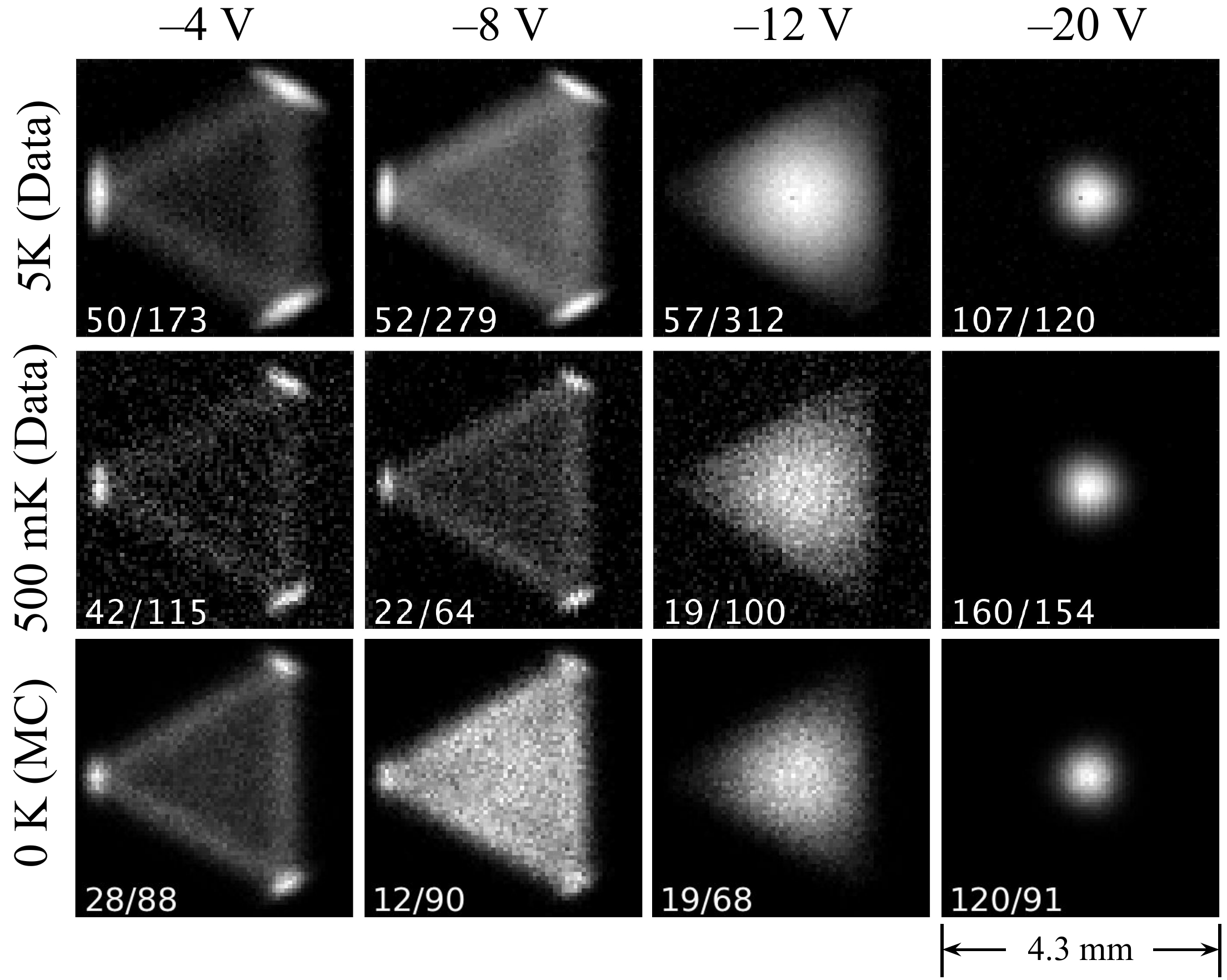}
    \caption{Two-dimensional electron collection density patterns for \Lv\ Si as a function of bias voltage. \textbf{First and second row}: Data recorded at 5\,K and 500\,mK. \textbf{Third row}: A simulation of the patterns in the zero-temperature limit. For the rows of data, each pixel is assigned a shade according to the measured pulse height, with white indicating larger pulses (and thus more electron collection). The white numbering shows the maximum pulse height (in meV) and the normalized integrated intensity (relative to $-12$\,V 500\,mK) for that panel. For the MC, a conversion was used such that peak intensity for the $-12$\,V simulation agrees with the $-12$\,V 500\,mK data.}
\label{fig:electrons_pattern}
\end{figure}

\section{Results}\label{sec:results}

The measured 2-D electron collection density patterns as a function of bias voltage can be seen in Figure~\ref{fig:electrons_pattern}. The results are shown for temperatures of 5\,K and 500\,mK, and measurements were performed at both temperatures for voltages between 4\,V and 20\,V, corresponding to field strengths of 10--50\,V/cm. These field strengths span the region in which the observed pattern changes the most, demonstrating that electrons at low field strength remain largely in the valley minima, and at high field strength undergo enough transitions to be concentrated along the field lines.



Large differences in electron collection are seen between 5\,K and 500\,mK, as shown in Figure~\ref{fig:electrons_pattern}.  We observe lower electron collection efficiency at 500\,mK, consistent with a large fraction of electron-hole pairs at this temperature either trapping or recombining before reaching the instrumented detector face. This temperature is low enough to allow for metastable over-charged states to form, but not low enough that these over-charged states become saturated during the course of a scan. In contrast, the 5\,K temperature data benefits from phonon-assisted transitions that make these low-energy bound states highly unstable, allowing the crystal to be more transparent regardless of the neutralization state. 



\begin{figure}[t]
    \centering
	\includegraphics[width=0.25\textwidth]{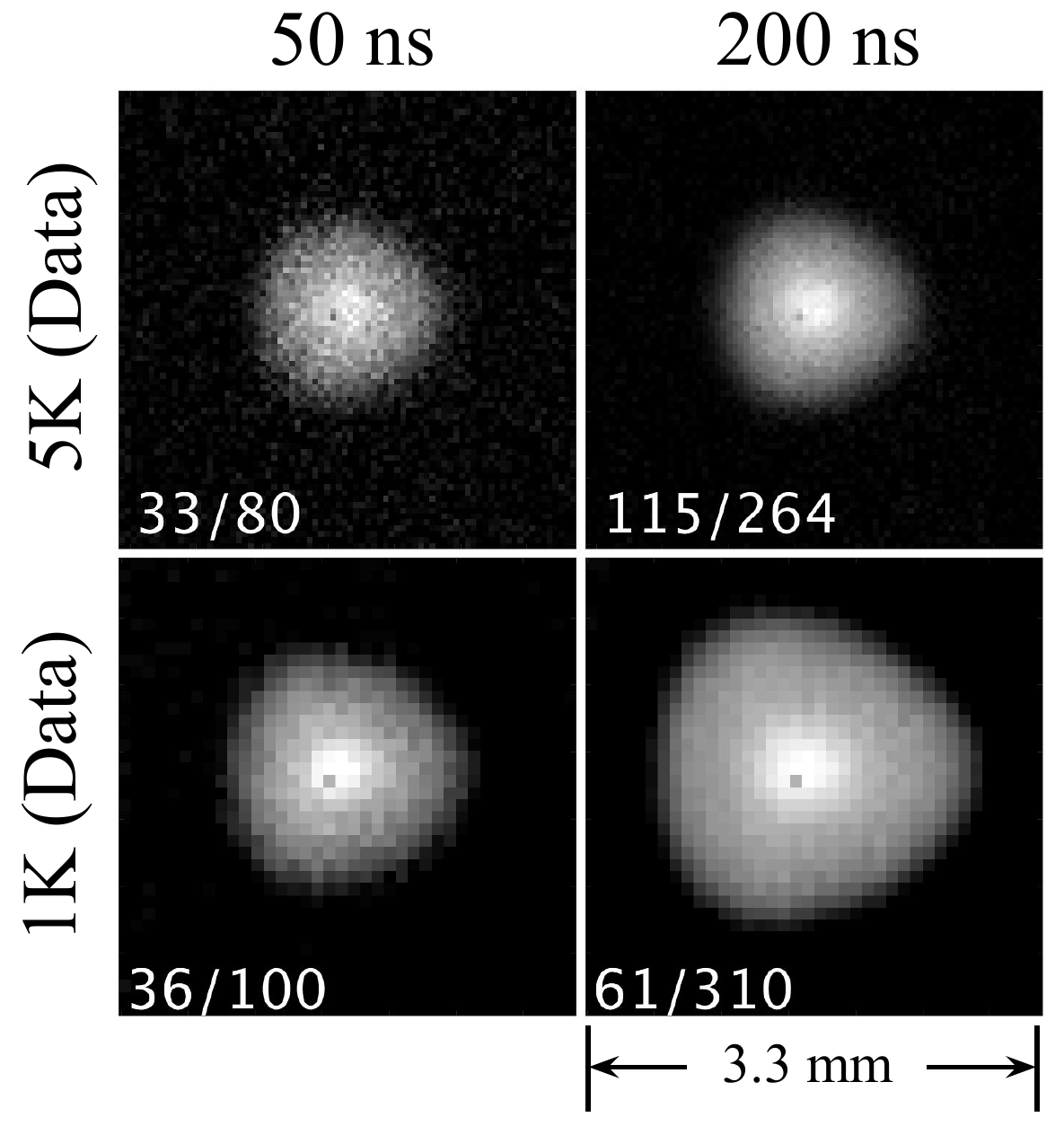}
    \caption{Two-dimensional hole collection density patterns at 5\,K and 1\,K for 6V crystal bias as a function of laser pulse width (50~ns or 200~ns). The white numbering shows the maximum pulse height (in meV) and the normalized integrated intensity (relative to $1$\,K 50\,ns). The more diffuse density patterns with the longer pulses are attributed to hole-hole repulsion.}
\label{fig:data_holes}
\end{figure}

Figure~\ref{fig:data_holes} shows a set of measured hole collection density images. All four scans were performed with at a bias voltage of +6\,V (corresponding to an electric field strength of 15\,V/cm). Each scan was performed with a laser pulse width of either 50\,ns or 200\,ns, and at a temperature of either 5\,K or 1\,K. All four of the hole density patterns show the same anisotropic shape. At the lower temperature, the size of the pattern is larger, which is attributed to a longer mean free path for the holes at lower temperatures. For scans performed with a higher laser pulse width, the size of the pattern is again seen to be larger, which we attribute to the effect of increased hole-hole repulsion in the initial hole cloud.



\begin{figure}[t]
\centering
\includegraphics[width=0.4\textwidth]{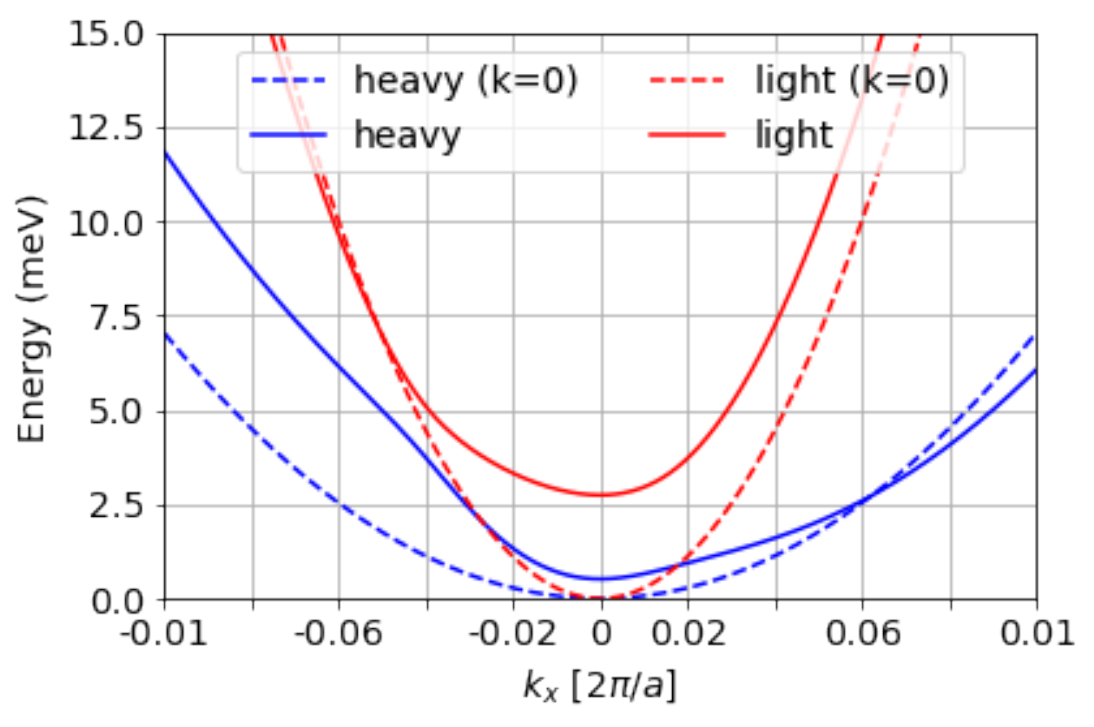}
\includegraphics[width=0.4\textwidth]{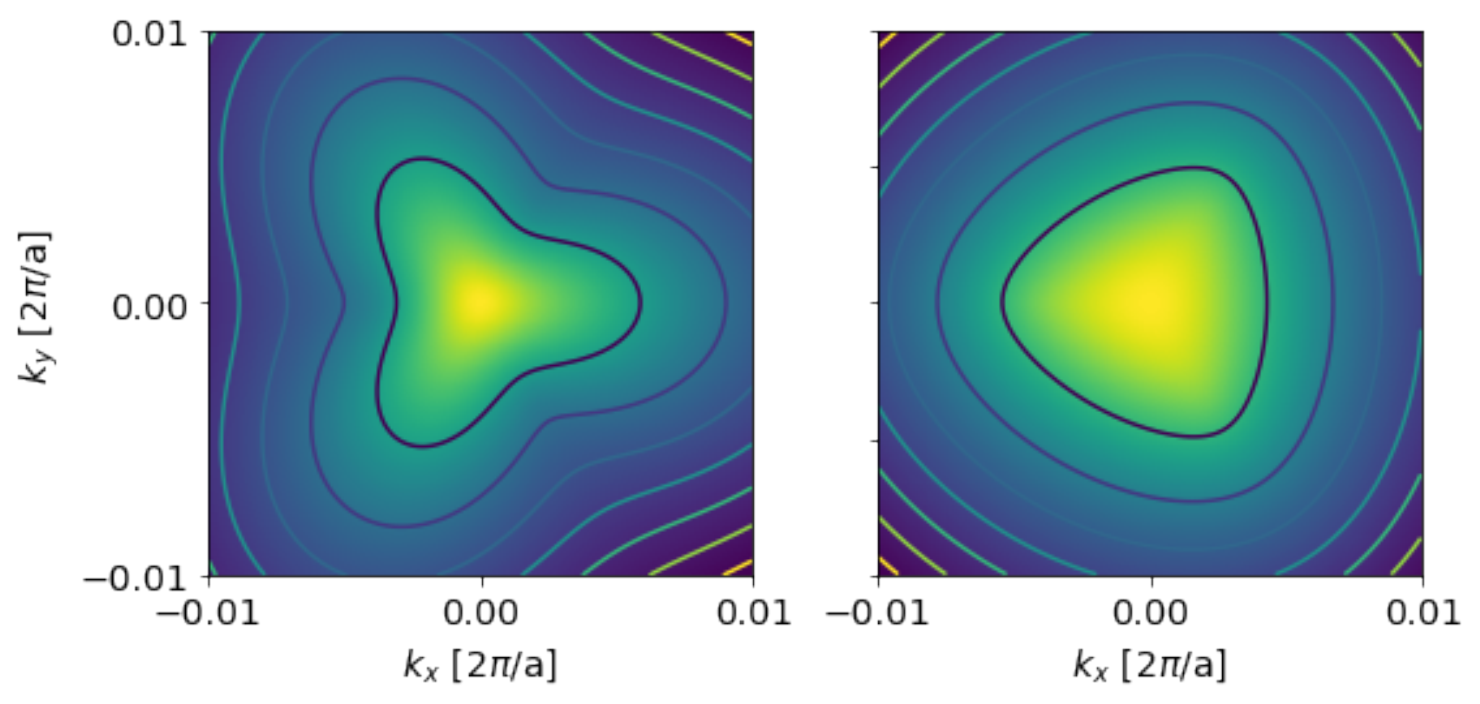}
\caption{Brillouin zone for the holes rotated so that $k_z$ is along the \Lv\ direction. \textbf{Top}: Dispersion relation at $k_y=0$ for holes drifting along the $z$ axis ($E_{heavy}=1$~meV) and stationary holes ($k_z=0$) near the band minimum. The equal energy surfaces for both the light and heavy holes as a function of crystal momentum become anisotropic for holes at non-zero drift velocity. For drifting holes, the heavy hole band will be preferentially filled by inter-band phonon emission. \textbf{Bottom}: Contours of constant kinetic energy for heavy (light) holes are shown on the left (right) where the center of the pattern shows contours for $\mathscr{E}\sim 3$~meV for the heavy holes and the mean energy of the contours is $\sim$10~meV, comparable to the mean hole energy for low electric field strength. The solid lines in the upper plot are cutlines through the lower plots for $k_y=0$.}
\label{fig:energy_surface}
\end{figure}

The anisotropy observed in the hole pattern, disregarding the aforementioned hole repulsion, is consistent with that expected for the heavy holes using the band constants from the literature\cite{DKK,ottaviani} shown in Figure~\ref{fig:energy_surface}, and we can enhance the effect of the anisotropy at low temperature and field strength by increasing the initial hole repulsion. This is to be expected for low temperatures; though the energy splitting between the heavy and light bands is small, there is a very high probability of emitting an optical phonon to transition from the light to heavy band, while the inverse transition is highly kinematically suppressed\cite{ottaviani,SundqvistThesis}. At higher field strengths (not shown), the hole pattern simply decreases in size with increasing electric field strength, while retaining the same shape. A Monte Carlo model for the hole propagation that generates this anisotropy is a subject of future work.

\section{Electron Intervalley Scattering Rate}\label{sec:MC}

To simulate cryogenic semiconductor detectors, we have developed an extension to GEANT for low-temperature charge and phonon transport (G4CMP)\cite{brandt}. We utilize the Herring-Vogt transform\cite{Jacoboni} to simplify simulation of the electron anisotropy, performing an isotropic Monte Carlo in the transformed momentum space. We have generalized this transform to an arbitrary number of valleys with user-defined longitudinal and transverse masses, allowing for the simulation of crystals with different Brillouin zone minima, and have used it to reproduce the results of our previous work\cite{Moffatt} as well as to simulate the data presented in this paper. A comparison of these simulations to data can be seen in Figure~\ref{fig:electrons_pattern}.

The primary motivation for the simulations shown in this paper was to use them to measure the intervalley scattering rate in Si as a function of voltage, as was done for Ge in \citet{Moffatt} and \citet{Broniatowski2014}. The mean free path of intervalley scattering can be deduced directly from measurements, but a model of the electron drift velocity obtained from the Monte Carlo is needed to convert these into scattering rates, which can then be compared to theory. 

For each electron collection image, the mean free path was determined by two methods. In the first method, a Gaussian profile was fit to the zero-scatter peaks to determine the fraction of electrons, denoted by $f_0$, which experienced zero intervalley scattering events. We then computed the mean free path, $\lambda$, as $\lambda = \eta/\ln(f_0^{-1})$, where $\eta\sim\SI{4}{\milli\meter}$ is the crystal thickness. We call this the ``integral method''. The integral method works well for data with appreciable zero-scatter populations, but fails when the mean free path is much smaller than the crystal thickness, in which case $f_0$ becomes too small to accurately measure. 

In the small $f_0$ limit, we can instead use the variance, $\sigma^2$, of the electron collection density pattern to determine the mean free path based on the amount of observed lateral diffusion, $\lambda\propto\sigma^2\eta^{-1}$. We call this the ``variance method''. The first method measures mean free path perpendicular to the crystal face, while the second measures the mean free path times a geometric factor which depends on the alignment of the electron valleys with respect to the crystal face. This geometric factor only depends on the Brillouin zone structure and the direction of the electric field relative to the lattice, but not on the electric field strength.


For each value of the electric field used in the experiment, simulated electron density patterns were generated with our Monte Carlo model. The same two techniques (integral \& variance) for estimating the mean free path were applied to both the measured data and to the Monte Carlo simulations. The intervalley scattering rate was used as a free fit parameter, and the closeness of fit was defined by how closely the mean free path calculated from the simulated electron density pattern reproduced the mean free path calculated from the 5~K data. This ensured that any geometric factors or unknown systematics introduced by these two techniques affected both the simulation and the data equally. Figure 5 shows, for each value of the electric field, the intervalley scattering rate of the Monte Carlo simulation which best matched the data.


\begin{figure}[t]
\centering
\includegraphics[width=0.45\textwidth,clip=true,trim=0 0 0 0]{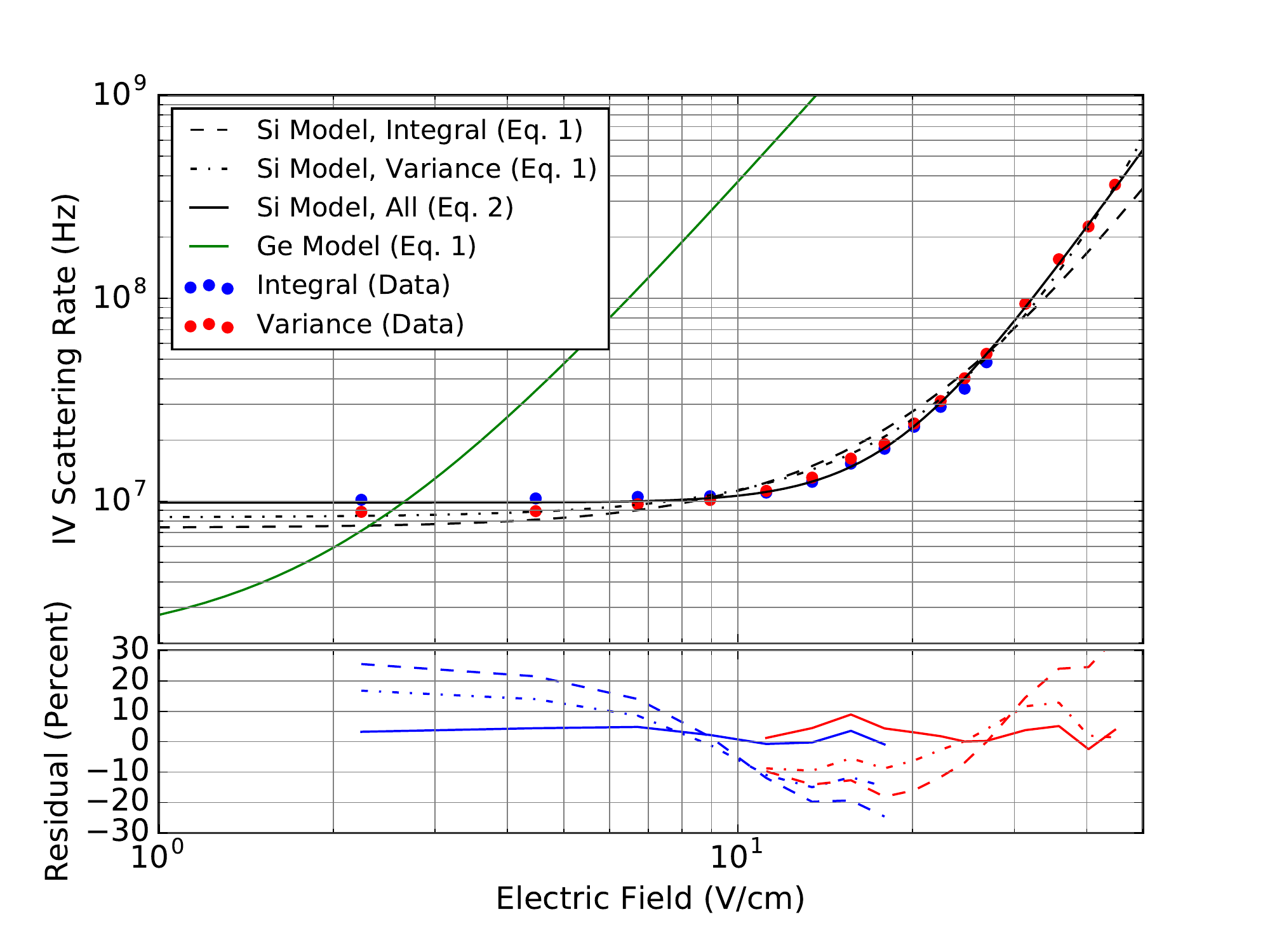}
\caption{Electron intervalley scattering rate as a function of electric field. The dots represent measurements using the integral (blue) and variance (red) methods respectively applied to the 5~K data, described further in the text. Also shown are fits of Equation~\ref{eq:IV1} to the red (dot-dashed line) and blue (dashed line) data points showing that this form cannot consistently fit the data. The fit of Equation~\ref{eq:IV2} (solid black) is a much better fit to the data across the entire range of electric field strengths, as shown by the residuals in the bottom panel. Also shown for comparison is the fit (green) of Equation~\ref{eq:IV1} to Ge IV scattering measurements from \citet{Moffatt}.}\label{fig:iv_scattering_rate}
\end{figure}


The rate measurements as a function of electric field~($E$) shown in Figure~\ref{fig:iv_scattering_rate} were fit to two different models. The first takes the form 
\begin{equation}\label{eq:IV1}
\Gamma_{IV} = \Gamma_0 \left[E_0^2+E^2\right]^{a/2}
\end{equation}
with best-fit parameters $\Gamma_0=\SI{3.5e-20}{\hertz(\meter/\volt)}^{\alpha}$, $E_0=\SI{3395}{\volt/\m}$, and $\alpha=7.47$. This is the same model that was used to fit the Ge measurements in~\citet{Moffatt}. It implicitly assumes that the limiting rates at low and high energy add in a nonlinear manner, implying that there is some correlation between intervalley scattering processes. This is non-physical, but was a good fit to Ge due to the rapid onset of phonon-assisted transitions and the smaller contribution to the rate from impurity scattering\footnote{The model used in~\citet{Moffatt} was not provided in that paper, but it is known through internal communication.}. However, if we assume that these processes are uncorrelated and add linearly as\cite{kurinsky}
\begin{equation}\label{eq:IV2}
\Gamma_{IV}=\Gamma_{I}+\Gamma_{Ph}\approx \Gamma_0+\Gamma_1 E^\alpha
\end{equation}
we obtain a much better fit to the entire range of electric fields in Figure~\ref{fig:iv_scattering_rate}. The best-fit parameters for this model are $\Gamma_0=\SI{9.8e6}{\hertz}$, $\Gamma_1=\SI{3.11e-7}{\hertz(\meter/\volt)}^{\alpha}$, and $\alpha=4.02$.

\begin{figure}[t]
\centering
\includegraphics[width=0.45\textwidth,clip=true, trim=40 30 70 20]{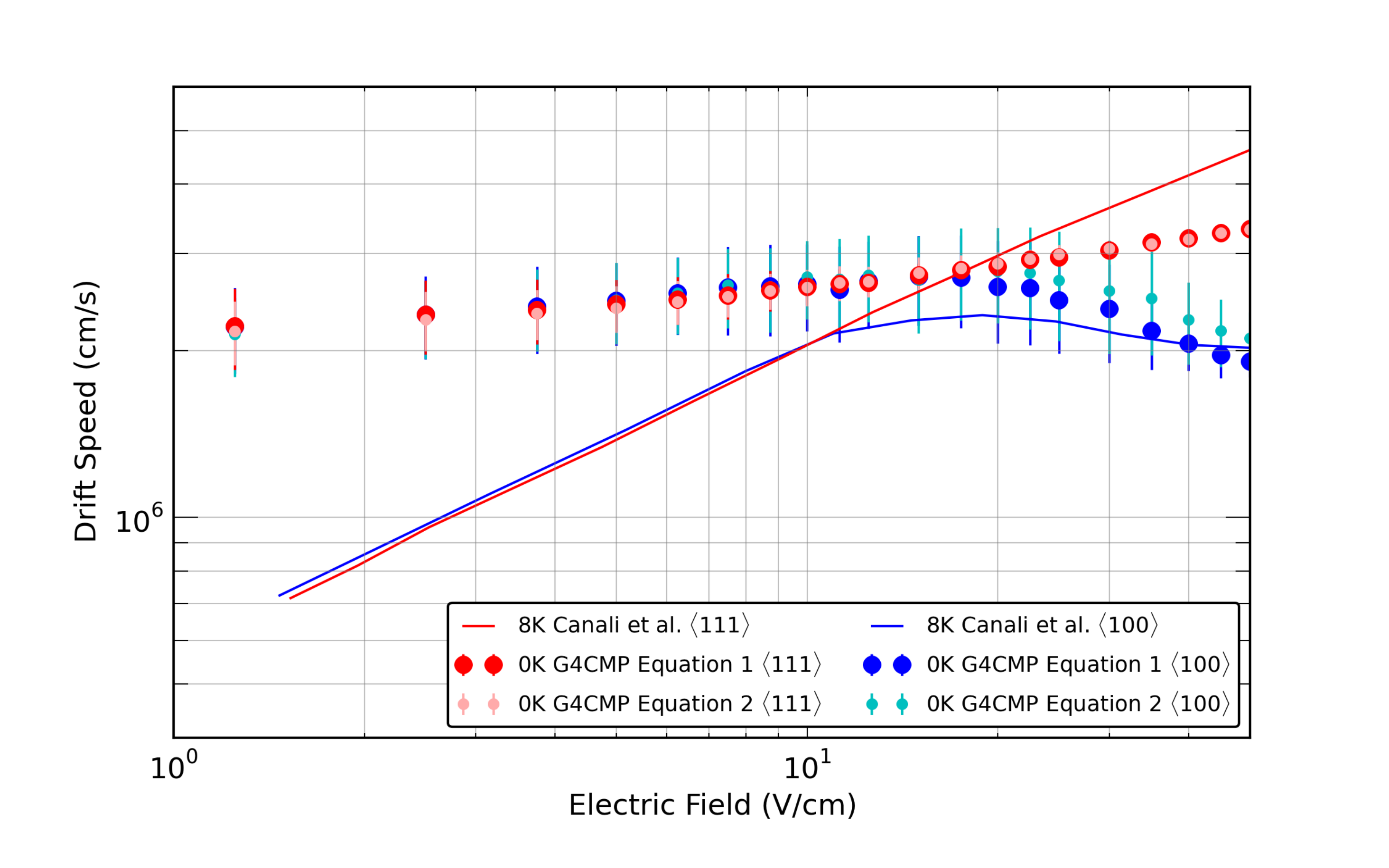}
\caption{Electron drift velocity as a function of electric field extracted from \citet{canali} compared to our simulations using G4CMP, both for the crystal orientation studied in this paper as well as the \Xv\ orientation. Some discrepancy is expected due to the differing temperatures.}
\label{fig:drift_velocity}
\end{figure}

To further validate our Monte Carlo, we compared the mean electron drift velocity from our simulations to the measurements performed by \citet{canali}, as shown in Figure~\ref{fig:drift_velocity}. Some discrepancy is expected, since the data were obtained at 8\,K and the simulation assumes a temperature of 0\,K. In particular, we expect a higher drift velocity at 0\,K for a given field strength. We see that the 0\,K drift velocity is higher at lower field strength, as expected, but at higher field strengths the G4CMP velocity curve for the \Lv\ orientation is slightly lower than the data. This discrepancy is likely due to our use of the Herring-Vogt transform to simulate electron-phonon scattering, which conserves energy but does not strictly conserve momentum. This is sufficient for our purposes, given that our simulations are concerned with spatial distribution rather than temporal charge properties. 

It should be noted that there are no drift velocity measurements to compare our simulations to at our operating temperature, so it is not straightforward for us to re-calibrate our Monte Carlo to data as was done in the past for Ge. This discrepancy does, however, suggest that we may be underestimating the scattering rate at higher field strengths. Direct measurements of these rates are underway in an effort to reduce this discrepancy.

\section{Discussion}

In this paper, we presented charge transport measurements in Si for electric fields strengths up to 50\,V/cm, demonstrating that the anisotropy in electron transport is retained up to $\sim$30\,V/cm for a 4~mm thick high-purity crystal. The anisotropic effects remain important up to much higher field strengths in Si than in Ge\cite{Moffatt}, which can be attributed to the higher optical phonon energies required for intervalley scattering in Si as compared to Ge\cite{Jacoboni,kurinsky}. The asymptotic field-independent impurity scattering rate at low electric field is higher than the rate measured for Ge by an order of magnitude\cite{Moffatt,Broniatowski2014}. A comparison between the data in this paper and the model fit from \citet{Moffatt} can be seen in Figure~\ref{fig:iv_scattering_rate}.

The observation of a flat intervalley scattering rate below 10\,V/cm, and the good fit to a power law at high electric field, suggests a scattering model that is dominated at high field by intervalley phonon transitions, and at low field by neutral impurity scattering, which is a very weak function of carrier energy\cite{Sclar,SclarI,kurinsky}. This is opposed to charged impurity scattering, for which the rate rises exponentially for low energy carriers (due to reduced charge screening effects), which would produce a rise in intervalley scattering rate at lower electric fields\cite{SundqvistThesis,kurinsky,weinreich}. This is consistent with the conclusions drawn by~\citet{Broniatowski2014} regarding the nature of the scattering centers in Ge. Measurements of intervalley scattering rate down to 20\,K in Si show that this effect is also temperature independent at this temperature scale, despite the large change in mean carrier energy, supporting a rate which is relatively flat in carrier energy\cite{weinreich}.

The neutral impurity scattering theory is further supported by the fact that ultrapure Si crystals (including the one employed in this experiment) tend to have impurity densities 10--100 times higher than the Ge crystals used in previous work\cite{Moffatt}. This helps account for the higher low-field IV scattering rate found in Si as compared to Ge. As indicated in \citet{Broniatowski2014}, the connection between impurity concentration and scattering rate is not monotonic. The rate of neutral impurity intervalley scattering depends both the number density of neutral impurity sites $n_{I}$ and the trapping energy $E_{T}$ as
$\Gamma_{IV}\propto n_{I}/E_{T}$\cite{kurinsky}. 
This is similar to that predicted by the model of neutral impurity electron exchange suggested in \citet{weinreich}. Work is ongoing to integrate this physical model into our detector Monte Carlo in order to verify that intervalley scattering can be modeled as scattering off of neutral impurity sites in low electric fields and as phonon-induced transitions in high electric fields\cite{Sclar,ridley1999}.

We have begun a program to extend these measurements to higher electric fields, where we expect impact ionization to start to play a larger role in electron transport, and we will use these data to help extend the Monte Carlo model to higher carrier energies. As part of this work we will continue to refine our models of charge collection and charge repulsion as a function of temperature, electric field, crystal orientation, and laser intensity to improve the quality of our simulations and to match the observed repulsion effects discussed in \citet{Moffatt}.

\acknowledgements{We thank Alexandre Broniatowski for discussions of intervalley scattering models developed for this paper. This work was supported in part by the U.S. Department
of Energy and by the National Science Foundation.}

\bibliographystyle{aipnum4-1}
\bibliography{APL_Si_Transport}

\end{document}